\documentclass[a4,12pt,epsf]{article}
\usepackage{amsmath,amssymb}
\usepackage{bm}
\usepackage{graphicx}
\usepackage{color}
\usepackage{wrapfig}
\begin{document}
\begin{center}{\Large \bf
Towards the minimal seesaw model
\vspace{6pt}\\
for the prediction of neutrino CP violation}
\end{center}
\begin{center}
Yusuke Shimizu$^a$
\footnote{yu-shimizu@hiroshima-u.ac.jp}
,
Kenta Takagi$^a$
\footnote{takagi-kenta@hiroshima-u.ac.jp}
,
Morimitsu Tanimoto$^b$
\footnote{tanimoto@muse.sc.niigata-u.ac.jp}
\vspace{6pt}\\
$^a$
{\it
Graduate School of Science
Hiroshima University,
Higashi-Hiroshima 739-8526, Japan}
\vspace{6pt}\\
$^b$
{\it
Department of Physics,
Niigata University,
Niigata 950-2181, Japan}
\end{center}
\begin{abstract}
We discuss the minimal seesaw model for the Dirac CP violating phase of the lepton mixing matrix.
We introduce two right-handed Majorana neutrinos
and obtain several textures of the tri-maximal lepton mixing matrices.
Moreover, we discuss the observed baryon asymmetry of the universe through the leptogenesis mechanism.
As the result, we obtain the specific model which predicts the negative sign of maximal Dirac CP violating phase
and normal hierarchy of neutrino masses.
\end{abstract}
\section{Our minimal seesaw model}
The remarkable developments in the neutrino oscillation experiments
fuel our expectations for the future discovery of CP violation in the lepton sector.
Indeed, recent T2K data strongly indicate the CP violation\,\cite{T2K}.
In order to discuss the theoretical aspects of the CP violation in the lepton sector,
we investigate the minimal seesaw model via the CP violation and baryon asymmetry of the universe (BAU).
Here, we briefly explain how to build our minimal seesaw model.
\begin{itemize}
  \item
  The minimal seesaw model includes two heavy right-handed Majorana neutrinos and three left-handed neutrinos
  in Type I seesaw\,\cite{Shimizu:2012ry}.
  We take both the charged lepton and right-handed Majorana neutrino mass matrix $M_R$ to be real diagonal.
  $M_R$ and the Dirac neutrino mass matrix $M_D$ are generally written as
\end{itemize}
\begin{align}
  M_R=-M_2
  \begin{pmatrix}
    p^{-1} & 0 \\
    0 & 1
  \end{pmatrix},\quad
  M_D=
  \begin{pmatrix}
    a & d \\
    b & e \\
    c & f
  \end{pmatrix},\quad
  p=M_2/M_1
\end{align}
\begin{itemize}
  \item[]
  the neutrino mass matrix is obtained by the Type I seesaw:
\end{itemize}
\begin{align}
  M_{\nu }=-M_DM_R^{-1}M_D^T=\frac{1}{M_0}
    \begin{pmatrix}
      a^2p+d^2 & abp+de & acp+df \\
      abp+de & b^2p+e^2 & bcp+ef \\
      acp+df & bcp+ef & c^2p+f^2
    \end{pmatrix}.
\end{align}
\begin{itemize}
  \item
  We consider the lepton mixing matrix in the two frameworks of tri-maximal mixing, ${\rm TM_1}$ and ${\rm TM_2}$
  which are derived from additional rotation of 2-3 and 1-3 plane
  to the tri-bi-maximal lepton mixing\,\cite{Harrison:2002er,Harrison:2002kp} respectively.
  The following textures of Dirac neutrino mass matrices realize the tri-maximal lepton mixing:
\end{itemize}
\begin{align}
  M_D=
  \begin{pmatrix}
    \frac{b+c}{2} & \frac{e+f}{2} \\
    b & e \\
    c & f
  \end{pmatrix},\quad
  \begin{pmatrix}
    -2b & \frac{e+f}{2} \\
    b & e \\
    b & f
  \end{pmatrix},\quad
  \begin{pmatrix}
    b & -(e+f) \\
    b & e \\
    b & f
  \end{pmatrix},
\end{align}
\begin{itemize}
  \item[]
  where they realize ${\rm TM_1}$ for normal hierarchy (NH),
  ${\rm TM_1}$ for inverted hierarchy (IH) and
  ${\rm TM_2}$ for either NH or IH of neutrino masses respectively.
\end{itemize}
\begin{itemize}
  \item
  We can discuss the BAU through the leptogenesis mechanism\,\cite{Fukugita:1986hr} in the decay of lighter right-handed neutrino $M_1$
  only for the ${\rm TM_1}$ with NH texture
  since only this texture produces the finite interference term between tree and 1 loop diagrams of the $M_1$ decay.
\end{itemize}
Therefore, we focus on the texture of ${\rm TM_1}$ with NH in the following.
In order to minimize our model, we impose zero in this texture.
The following three types of Dirac neutrino mass matrices are possible.
\begin{align}
  M_D^{\rm I}=
  \begin{pmatrix}
    0 & \frac{e+f}{2} \\
    b & e \\
    -b & f
  \end{pmatrix},\quad
  M_D^{\rm II}=
  \begin{pmatrix}
    \frac{b}{2} & \frac{e+f}{2} \\
    b & e \\
    0 & f
  \end{pmatrix},\quad
  M_D^{\rm III}=
  \begin{pmatrix}
    \frac{c}{2} & \frac{e+f}{2} \\
    0 & e \\
    c & f
  \end{pmatrix}
\end{align}
\section{Numerical analysis}
We discuss the correlation between the predicted CP violating phase $\delta_{CP}$
and the BAU through the leptogenesis.
Our analysis about the leptogenesis mainly follows a simple framework\,\cite{Giudice:2003jh}
which is valid under the condition, $M_1\ll M_2$ and $M_1\ll10^{14}$[GeV].

\begin{figure}[t]
	\begin{tabular}{ll}
		\begin{minipage}{0.475\linewidth}
			\begin{center}
\includegraphics[{width=\linewidth}]{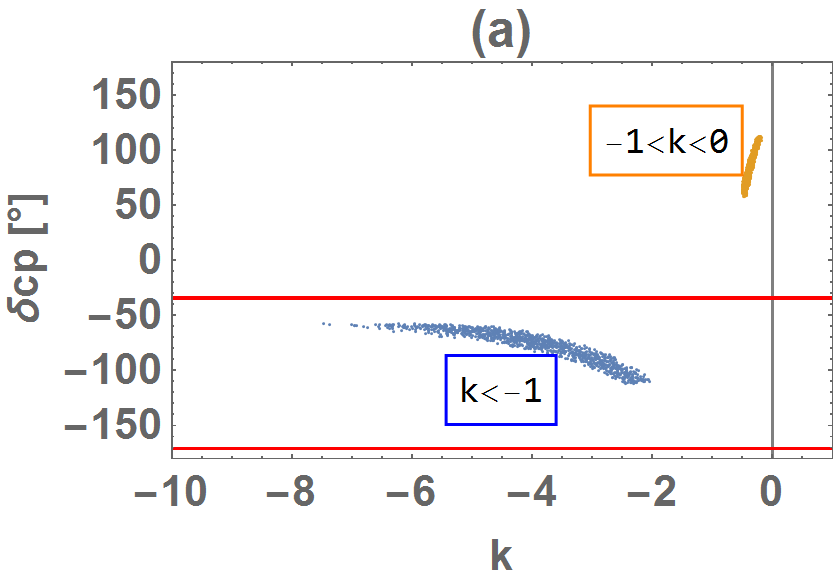}
			\end{center}
		\end{minipage}
		\begin{minipage}{0.475\linewidth}
			\begin{center}
\includegraphics[{width=\linewidth}]{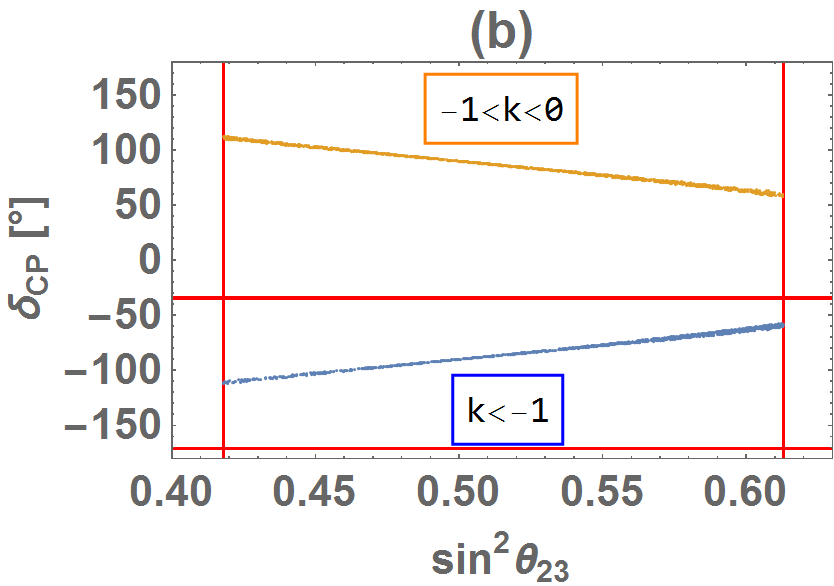}
			\end{center}
		\end{minipage}
	\end{tabular}
	\caption{Predictions in case I, where
		the blue  and orange dots denote the region of $k<-1$
		and $-1<k<0$. The  red lines for $\sin^2\theta_{23}$
    and $\delta_{CP}$ denote the experimental bounds of $3\sigma$ (global analyses)
    and $2\sigma$ (T2K) ranges, respectively:
		  (a)  $\delta_{CP}$ versus $k$,
		(b)   $\delta_{CP}$ versus $\sin^2\theta_{23}$.
	}
	\label{fig:case1deltaCP}
\end{figure}

Here, we discuss the numerical results of ${\rm TM}_1$ with NH.
We use the recent neutrino oscillation data from NuFIT 3.2 (2018)\,\cite{Esteban:2016qun}
for the input data.
According to this global experimental data, the numerical results from
$M_D^{\rm II}$ and $M_D^{\rm III}$ are excluded from $3\sigma$ C.L..
Therefore, we only show the results of $M_D^{\rm I}$  in Figure\,\ref{fig:case1deltaCP}.
The results reflect the constraint from not only the recent neutrino oscillation data but also the observed BAU,
$\eta_B\simeq[5.8,6.6]\times10^{-10}$ $95\%$C.L.\,\cite{Olive:2016xmw}.
These predictions are calculated in the case of $M_2=10^{14}$GeV.
But the correlations in Figure\,\ref{fig:case1deltaCP} are independent of $M_2$.
We note that the ratio of right-handed neutrino masses $p=M_2/M_1$ is allowed ,roughly speaking,
within $p=[200\sim300]$ ($p=[2000\sim3000]$) for $M_2=10^{14}$[GeV] ($M_2=10^{15}$[GeV]).

Let's discuss the left panel of Figure\,\ref{fig:case1deltaCP}.
We show the $k=e/f$ dependence of the predicted $\delta_{CP}$ by inputting the observed BAU.
It is remarked that $\delta_{CP}$ is predicted to be negative for $k<-1$ while  it is positive for $-1<k<0$.
The negative and maximal CP violation $\delta_{CP}\sim-\pi/2$ is realized around $k\sim-3$.

In the right panel of Figure\,\ref{fig:case1deltaCP},
we show the predicted correlation between $\delta_{CP}$ versus $\sin^2 \theta_{23}$.
It indicates an important feature of this model:
The maximal CP violation and mixing angle
$(\delta_{CP},\theta_{23})=(-\pi/2,\pi/4)$ can be realized for $k<-1$ simultaneously,
which is favored if we take account of the current data and future prospects.

\section{Summary and discussions}
We have studied the correlation between the CP violating phase $\delta_{CP}$
and the observed BAU in the minimal seesaw model,
where two right-handed Majorana neutrinos are assumed.
We have also taken the tri-maximal mixing pattern for the neutrino flavor ($\rm TM_1$ or $\rm TM_2$) in the diagonal basis of both the charged lepton and right-handed Majorana neutrino mass matrices.
We have found the clear correlation between the CP violating phase $\delta_{CP}$ and  BAU for $\rm TM_1$ in NH of neutrino masses.
The parameter $k$ should be smaller than $-1$ in order to predict a negative $\delta_{CP}$,
which is indicated by the recent T2K data.
It is emphasized that our Dirac neutrino mass matrix predicts
the negative sign of $\delta_{CP}$ and the observed value of BAU
as far as we take $k<-1$ under the condition, $M_1\ll M_2$.
\vspace*{12pt}
\noindent
\\
{\bf Acknowledgement}
\vspace*{6pt}
\noindent
This work is supported by JSPS Grants-in-Aid for Scientific Research
 16J05332 (YS) and 15K05045, 16H00862 (MT).

The full references are available in
[arXiv:1709.02136] and
[arXiv:1711.03863].

\begin{thebibliography}{99}
\bibitem{T2K}
T2K report, http://t2k-experiment.org/2017/08/t2k-2017-cpv/, August 4, 2017.

\bibitem{Shimizu:2012ry}
  Y.~Shimizu, R.~Takahashi and M.~Tanimoto,
  PTEP {\bf 2013} (2013) no.6, 063B02
  [arXiv:1212.5913 [hep-ph]].

\bibitem{Harrison:2002er}
  P.~F.~Harrison, D.~H.~Perkins, W.~G.~Scott,
  Phys.\ Lett.\ B {\bf 530 } (2002)  167
  [hep-ph/0202074].


\bibitem{Harrison:2002kp}
  P.~F.~Harrison, W.~G.~Scott,
  Phys.\ Lett.\ B {\bf 535 } (2002)  163-169
  [hep-ph/0203209].

\bibitem{Fukugita:1986hr}
  M.~Fukugita and T.~Yanagida,
  Phys.\ Lett.\ B {\bf 174} (1986) 45.

  \bibitem{Giudice:2003jh}
  G.~F.~Giudice, A.~Notari, M.~Raidal, A.~Riotto and A.~Strumia,
  Nucl.\ Phys.\ B {\bf 685} (2004) 89
  [hep-ph/0310123].

  \bibitem{Esteban:2016qun}
  NuFIT 3.2 (2018), www.nu-fit.org/

  \bibitem{Olive:2016xmw}
  C.~Patrignani {\it et al.} [Particle Data Group],
  Chin.\ Phys.\ C {\bf 40} (2016) no.10,  100001.

\end{thebibliography}
\end{document}